\documentclass{webofc}
\usepackage[varg]{txfonts}   % Web of Conferences font
\def \arcsec{\hbox{$^{\prime\prime}$}\,}
\def \arcmin{\hbox{$^{\prime}$}\,}

\def \micb{$\mu$m }
\def \tw{$^{12}$CO }
\def \deg{$^\circ$}
\begin{document}
\title{NOEMA complementarity with NIKA2}
%
% subtitle is optional
%
\subtitle{To appear in the proceedings of the international conference entitled mm Universe @ NIKA2, Grenoble (France), June 2019, EPJ Web of conferences }

\author{\firstname{Charlène} \lastname{Lefèvre}\inst{1}\fnsep\thanks{\email{lefevre@iram.fr}} \and 
	\firstname{Carsten} \lastname{Kramer}\inst{1} 
	\and
	\firstname{Roberto} \lastname{Neri}\inst{1}
			\and
	\firstname{Stefano} \lastname{Berta}\inst{1}
		\and
	\firstname{Karl} \lastname{Schuster}\inst{1}
        % etc.
}

\institute{Institut de RadioAstronomie Millimétrique (IRAM), 300 rue de la Piscine, 38406 Saint Martin d'Hères
%\and
%           the second here 
%\and
%           Last address
          }

%\abstract{%
%	IRAM operates two observatories - the 30-meter Telescope with its new NIKA2 camera on Pico Veleta in Spain and NOEMA, an interferometer of ten 15-meter antennas on Plateau de Bure in France. Both {observatories} allow to observe at millimeter wavelengths. We aim at comparing them and discuss the complementarity between NOEMA and NIKA2 and their role at the cutting edge of research in astronomy. In particular, we will review the possible synergies from nearby star forming regions to high redshift galaxies at cosmological distances.
%}

\abstract{%
	IRAM operates two observatories - the 30-meter Telescope on Pico Veleta in Spain and NOEMA, an interferometer of ten 15-meter antennas on Plateau de Bure in France. Both {observatories} allow to observe at millimeter wavelengths. Here, we aim at discussing the complementary between continuum observations with NOEMA and NIKA2 at the 30m and their role at the cutting edge of research in astronomy. In particular, we will review possible synergies of continuum studies from nearby star forming regions to high redshift galaxies at cosmological distances.
}

\maketitle
\section{Introduction}
\label{intro}

 IRAM telescopes, the Northern Extended Millimeter Array (NOEMA) and the 30m Telescope, operate in the four atmospheric windows {at at 0.8, 1.0, 2.0 and 3.0 mm}. In this wavebands, it is possible to detect simultaneously thousands of spectral lines as well as thermal continuum spectra from cold dust.  {NOEMA and the 30m Telescope, combined, offer the unique opportunity to study the sky on spatial scales over more than four orders of magnitude, between few 0.1\arcsec and few degrees.} NOEMA is more sensitive with higher angular resolution {but smaller FoV} compared to the 30m Telescope. {We aim at discussing the capabilities of both observatories for broad band continuum observations.} The 30m Telescope hosts NIKA2, {plus two heterodyne frontends, EMIR and HERA}.
 %, while NOEMA receivers fill the space in the cabin of each 15-meter antenna. 
 NIKA2 is an array made of KIDs dedicated to continuum detection {\cite{2018Adam}} that covers a fixed frequency range corresponding to the 2\,mm atmospheric window and about one third of the 1\,mm atmospheric window (see table \ref{tab-1}). On the other hand, NOEMA receivers can be tuned in the {four} atmospheric bands{, like EMIR,} and offer a frequency coverage of $\sim$\,31\,GHz corresponding to 2 polarizations made of two separated side bands of 7.744 GHz each. NOEMA heterodyne receivers are combined to a powerful correlator (PolyFiX) to retrieve simultaneously continuum and spectral information. The complementary between the two facilities and the two instruments make a unique opportunity to explore the millimeter Universe.
 
 %Also, with NOEMA the image plane is partly sampled and large scale structures are inevitably filtered out, unlike the 30-meter. 

\section{NOEMA and NIKA2}
\label{comparison}

In order to compare their capabilities, we provide below key numbers for NOEMA and NIKA2 single pointing continuum sensitivities and FoV. The following values are not taking into account the fast scanning capabilities of NIKA2 compared to NOEMA but we will discuss in Sect.~\ref{point_contamination} NOEMA capabilities for mosaics.\\

\noindent NOEMA point source continuum sensitivity with 10 antennas is the following:
\begin{itemize}
	\item at 2\,mm (150\,GHz) in a FoV of 33\arcsec in 1\,h on source with PolyFiX{\footnote{{To cover the same bandwidth than NIKA2, 4 NOEMA setups would be needed, meaning a factor 2 in terms of sensitivity with respect to the numbers presented here. Nevertheless, this is not mandatory for all scientific goals.}} ($\sim$\,31\,GHz}):
\end{itemize}
\begin{equation}
\label{eq:noema_sensitivity_1}
\begin{array}{l}
26\,\mathrm{Jy/K}\,.\,\dfrac{{117}}{{0.90}\,.\,\sqrt{90 \times 15.5\,10^9 \times 2 \times 3600} } \simeq {34}\, \mathrm{\mu Jy/beam}
\end{array}
\end{equation}

\begin{itemize}
	\item at 1.15\,mm (260\,GHz) in a FoV of 19\arcsec:
\end{itemize}
\begin{equation}
\label{eq:noema_sensitivity_2}
\begin{array}{l}
33\,\mathrm{Jy/K}\,.\,\dfrac{{209}}{0.80\,.\,\sqrt{90 \times 15.5\,10^9 \times 2 \times 3600} } \simeq {86}\, \mathrm{\mu Jy/beam}
\end{array}
\end{equation}

\noindent These numbers can be compared to NIKA2 Time Estimator using the following formula \cite{2018Ladjelate}:

\begin{equation}
\begin{array}{l}
t_{on-source} = \,\left(\dfrac{\mathrm{NEFD_0}\cdot e^{\tau/\mathrm{sin(el)}}}{\sigma}\right)^2 \times \left(1 + \dfrac{\Delta_x\cdot\Delta_y}{f_{pix}\cdot A_\mathrm{FoV}}\right),
\end{array}
\end{equation}
\noindent where $\Delta_x\cdot\Delta_y = A_\mathrm{FoV}$, {$f_{pix}$ is the fraction of good pixels in the array} and observing conditions are similar to Eq.\,\ref{eq:noema_sensitivity_1}--\ref{eq:noema_sensitivity_2} with a typical elevation (el) of 25\deg\,and an opacity ($\tau$) corresponding to a precipitable water vapor (pwv) amount of {3\,mm}. $\mathrm{NEFD_0}$ are taken from \cite{2018Adam}. After 1\,h on source, NIKA2 single pointing gives a sensitivity $\sigma$\,$\sim$\,{0.195}\,mJy/beam at 2\,mm and $\sigma$\,$\sim$\,1.15\,mJy/beam at 1.15\,mm ({a factor {5.7} and {13.4} relative to NOEMA, respectively}). {Sensitivity being in favor of NOEMA$^1$, all {point} sources detected with NIKA2 can be observed at NOEMA with a high dynamic range.} Nevertheless, NOEMA is designed to observe at high angular resolution with beam sizes depending on the array configuration (see table 1). Also, the NIKA2 FoV is more than a factor 100 larger than {that} of NOEMA with a fast mapping speed, and depending on the science goal the NOEMA advantage may shrink (see sect. \ref{point_contamination}). Hence, the complementarity should take advantage of the difference of scales.

\begin{table}[h!]
	\centering
	\caption{NOEMA and NIKA2 specifications for a single pointing. Numbers below do not take into account fast mapping capabilities of NIKA2. NOEMA beam sizes are indicated for the three array configurations currently available (extended: A, intermediate: C, compact: D). $\sigma$ is the point source continuum sensitivity obtained {after one hour on source}. More details are given in Sect.~\ref{comparison}.}
	\label{tab-1}       % Give a unique label
	% For LaTeX tables you can use
	\begin{tabular}{|c|c|c|c|c|c|c|c|}
		\hline
		& $\Delta\nu$ & Frequency &FoV & \multicolumn{3}{c|}{Beam size (\arcsec)} & $\sigma$\\
		& GHz & range (GHz) &  & A & C & D & $\mu$Jy/beam \\
		\hline
		NOEMA 2\,mm (150\,GHz)& 31$^1$ &  127 -- 183 & 33\arcsec & 0.6 & 1.3 & 2.6 & 34 \\
		\hline
		NIKA2 2\,mm & 50 & 125 -- 175 &  6.5\arcmin & \multicolumn{3}{c|}{17.7} & 195\\
		\hline	
		\hline
		NOEMA 1\,mm (260\,GHz) & 31$^1$ & 196 -- 276 & 19\arcsec & 0.35 & 0.75 & 1.5 & 86\\
		\hline
		NIKA2 1\,mm & 50 & 230 -- 280 & 6.5\arcmin & \multicolumn{3}{c|}{11.2}& 1150 \\
		\hline
		
	\end{tabular}
\end{table}
 
\section{NOEMA and NIKA2 Synergies}
\label{synergies}
High angular resolution reached by interferometers allow to explore inner parts of the sources but also to avoid confusion between different objects. In particular, NOEMA high angular resolution gives access {in more detail} to the molecular gas reservoir for star formation and cold dust far away in high-redshift galaxies, as well as nearby proto-stellar envelopes. In this Section, we will present a non-exhaustive list of synergies between NOEMA and NIKA2.

\subsection{Zooming in star-forming regions}
\label{zoom-SF}

\begin{figure}[h]
 \centering
 \sidecaption
 \includegraphics[width=9cm,clip]{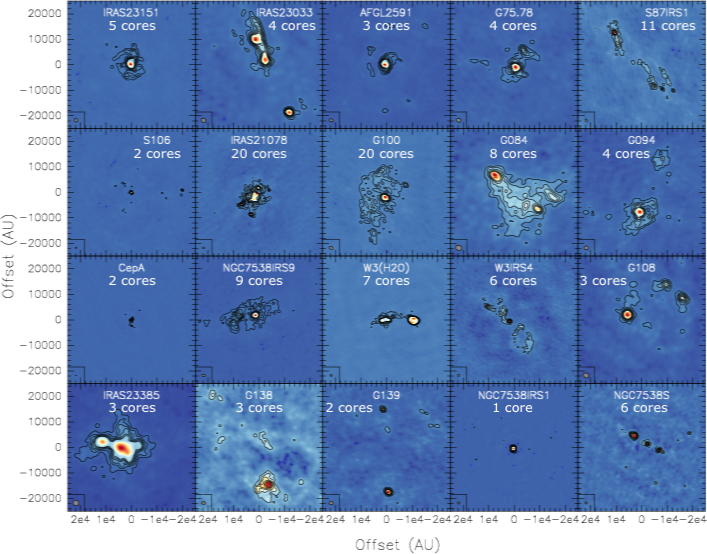}
\caption{1.37\,mm continuum images for the 20 high-mass star forming regions observed with NOEMA ($\sim$\,18\,h per region) in the frame of CORE \cite{2018Beuther}. Each of them was previously detected as a single continuum source in SCUBA-2 850\,\micb continuum data \cite{2008DiFrancesco} except G100, G084, and G108. Sources are labeled in each panel with the number of cores detected. Synthesized beams are shown at bottom-left. The contouring is in $5\sigma$ steps.}
 \label{fig-1}       % Give a unique label
 \end{figure}

High mass stars play a major role in interstellar medium (ISM) via their feedback and the release of heavy elements to the surrounding medium. They form mainly in proto-clusters embedded within their natal cores. Single dish telescopes cannot make the distinction between a single massive proto-stellar core or a deeply embedded proto-cluster containing several cores. Interferometers are required to reach sub-arcsec scale for high mass star forming regions located at several kpc. Knowing the degree of multiplicity of high mass stars is crucial to understand their formation processes from cores to stars \cite{2014Offner}. In this framework, the IRAM Large Program (LP) CORE zoomed in 20 high-mass star forming regions with high luminosities indicating that at least an 8\,M$_\odot$ star is forming \cite{2018Beuther}. SCUBA-2 data at 850\,\micb showed apparently monolithic continuum sources \cite{2008DiFrancesco} but NOEMA {combined with the 30m} revealed complex morphologies ranging from regions dominated by a single high-mass core to regions that fragment into up to 20 cores at 1.37\,mm with an angular resolution of 0.3\,-\,0.4\arcsec (see Fig.~\ref{fig-1}). The fragmentation diversity between regions with few or only one massive core compared to those with many low-mass cores may be explained by different formation mechanisms that coexist, or even interplay, or by differences in the initial density profiles and/or magnetic field properties of the parent gas clump. Thanks to sample selection {of CORE LP}, evolutionary stage and turbulence can be ruled out as the main contribution for fragmentation diversity. Since the nearest neighbor separation peaks close to NOEMA's angular resolution, it is likely that further fragmentation takes place on smaller spatial scales, that could be investigated {even} at higher resolution. Indeed, an apparently monolithic continuum source showed fragmentation signatures down to 0.2\arcsec($\sim$\,500\,AU) relying on kinematic analysis of HCN at 0.8\,mm \cite{2013Beuther}. NIKA2 with its large FoV and fast mapping speed is well suited {to discover} high-mass candidates inside star forming regions to be followed up with NOEMA. Moreover, NOEMA will double its baseline lengths in 2020, pushing towards higher angular resolution and making possible to observe {more distant} star forming regions.
 
 %For figure with sidecaption legend use syntax of figure

Combining single-dish and interferometric continuum data can also reveal a column density enhancement in a dense core which has not previously traced by single-dish observations alone. This was the conclusion of a study combining IRAM MAMBO-2 and the Alma Compact Array (ACA) data that found a centrally concentrated sub-solar mass starless core in the Taurus L1495 filamentary complex \cite{2019Tokuda}. Because this core is very dense, with a high deuterium fraction it indicates that it should be at the very end of its starless phase. Hence, it may be the precursor of a brown dwarf or a very low-mass star. IRAM Telescopes will help to understand the very low mass star formation. First, NIKA2 will provide a unique opportunity to look for pre-brown dwarf candidates (see proceeding by B. Ladjelate) {and to perform square-degree studies of Galactic clouds (e.g. the GASTON LP, see proceeding by Peretto et al.)}. In a second step, NIKA2 and NOEMA continuum data could be directly combined by attributing pseudo-visibilities to NIKA2 short spacings. 
%Complementary NOEMA observations will be essential to confirm the column density enhancement.
%\begin{figure}[h!]
%\begin{center}
%
%\end{center}
%\label{fig-1}       % Give a unique label
%\end{figure}
\subsection{Line and point source contamination}
\label{contamination}

\paragraph{Line contamination}
\label{line_contamination}
Because of its broad band, {emission in NIKA2 bands is not coming only from continuum but also from lines}. In particular, for nearby sources the \tw emission line at 230.538 GHz lies towards the low frequency edge of the 1.15\,mm NIKA2 band (see Table \ref{tab-1}). The typical contribution of \tw for galactic filaments was estimated to 1\,--\,3\% in NIKA studies \cite{2017Bracco,2018Rigby}. Line surveys have shown forest of lines in nearby clouds \cite{2019Fuente} and galaxies \cite{2015Aladro} suggesting that CO is likely the dominant but not the only contamination inside NIKA2 wavebands. Moreover, CO and water line emission are also detected in strongly lensed galaxies e.g. by ALMA at z\,$\sim$\,3.6 \cite{2019Yang-ALMA} and by NOEMA at z\,$\sim$\,6.5 \cite{2019Yang-NOEMA} suggesting that ancillary data may not be sufficient to properly estimate the line contamination to NIKA2 data. With its large frequency coverage, NOEMA can be used to do line studies of high-z sources, like done by EMIR \cite{2009Weiss}, and estimate line contamination.

\vspace{-0.3cm}
\paragraph{Point source contamination}
\label{point_contamination}

\begin{figure}[b!]
	\centering
	\begin{minipage}[b]{0.45\linewidth}
		\includegraphics[width=0.85\textwidth]{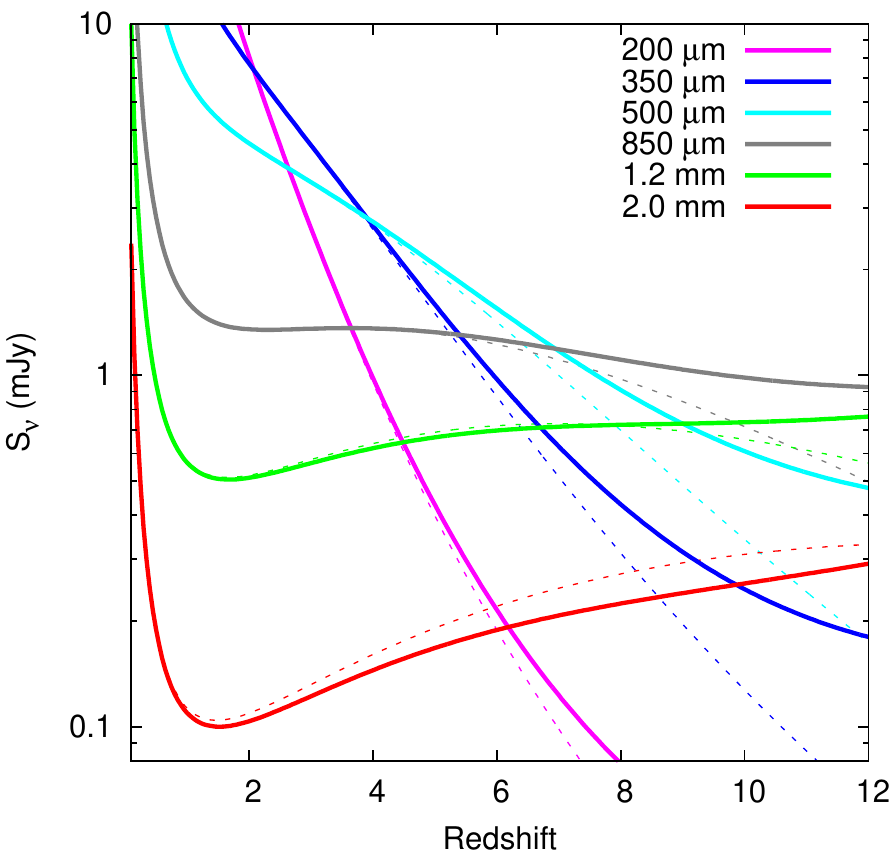}
	\end{minipage}
	\begin{minipage}[b]{0.45\linewidth}
		\includegraphics[width=0.85\textwidth]{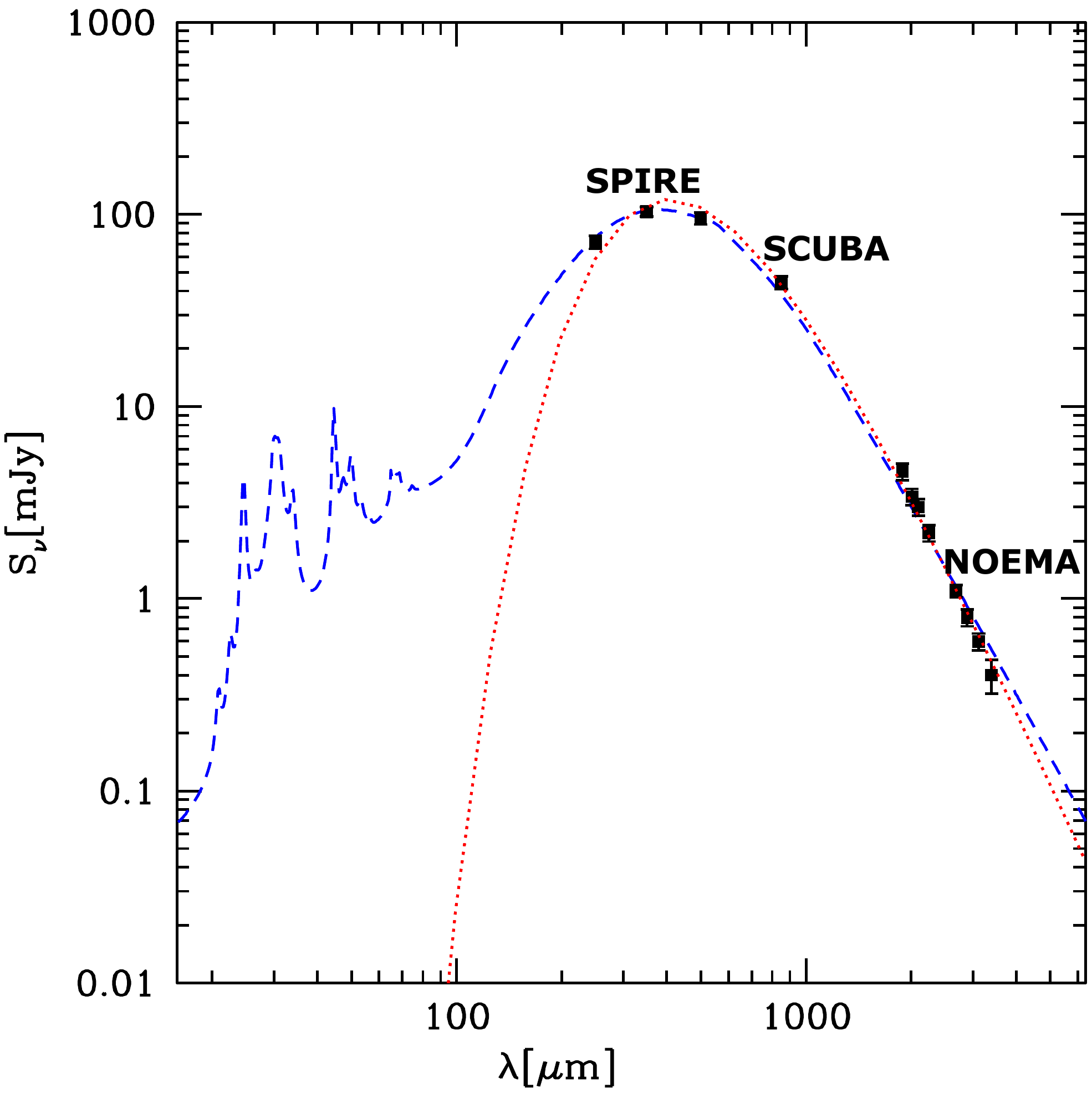}
	\end{minipage}
	\caption{{Left:} Redshift-dependent flux density of a
		galaxy with fixed dust mass of 10$^8$ M$_\odot$ and a dust emissivity index of
		$\beta$ = 1.5 from Staguhn et al. \cite{2014Staguhn}. {Right:} SED of a Herschel selected gravitational lens at z\,$\sim$\,3. The NOEMA part of the SED is measured with an accuracy better than 10\% in about one-hour on-source (Berta et al., in prep.).}
	\label{fig-2}       % Give a unique label
\end{figure}

Blind survey of distant galaxies, without any pre-selection of the targeted sources, represent a unique approach to understand the formation and evolution of dust across cosmic time and to determine the cosmic history of star formation from their gas content \cite{2014Madau}. With its high mapping speed, NIKA2 will detect hundreds of dust-obscured optically-faint galaxies. 
The flux density of galaxies at a redshift z\,>\,1.5 stop to decrease with distance at submm/mm wavelengths making high-redshift galaxies easy to detect compared to low-redshift ones (see Fig.~\ref{fig-2} left panel). Once candidates are detected, Deep Field surveys benefit from cross-correlations with catalogs at other wavelengths to build the Spectral Energy Distribution (SED). NOEMA with its 8 individual basebands gives a good sampling of the SED that can be extend down to the 3\,mm band (see Fig.~\ref{fig-2} right panel, and last but not least provide also the molecular gas emission \cite{2014Decarli, 2014Walter}. Accurate point source flux measurement thanks to SED also allow to remove point source contamination {to the SZ signal} (see Sect.~\ref{SZ} and F. Ruppin's proceeding). {Thanks to the large bandwidth of PolyFiX and increased sensitivity, NOEMA improved its mapping capabilities and can reach for a 60\deg~declination source in 2019 a sensitivity of:}
\begin{itemize}
	\item {0.2\,mJy/beam for 4\,arcmin$^2$ at 260\,GHz in 12.7\,h of telescope time\footnote{Overheads include calibrations, winter conditions (pwv=3\,mm) were used. The time needed will decrease by a factor 1.46 with the 12 antenna array in 2021.}\,($\sim$\,7.3\,h on source)}
	\item {0.05\,mJy/beam for 9\,arcmin$^2$ at 150\,GHz in 24.4\,h of telescope time ($\sim$\,14\,h on source)}
	\item {20\,$\mu$Jy/beam for 6 arcmin$^2$ at 100\,GHz in 12.8\,h  of telescope time ($\sim$\,7.3\,h on source)}
\end{itemize}
%this refers

\vspace{-0.6cm}
\subsection{Sunyaev-Zel'dovich effect}
\label{SZ}
Clusters of galaxies are the most massive structures in the early Universe. The Intra Cluster Medium (ICM) contains free hot electrons that interact with the Cosmic Microwave Background (CMB). This interaction when electrons are highly energetic due to their temperature is called the thermal Sunyev-Zel'dovich (tSZ) effect. In practice, it shifts the CMB photons toward higher energies producing a distorted spectrum. This effect is proportional to the number density of electrons, the thickness of the cluster along the line of sight, and the electron temperature. The NIKA2 LP SZ aims at better characterizing the thermodynamic properties of clusters of galaxies and the mass-scaling relation of the ICM with redshift (see proceeding by F. Mayet). Below we will discuss the complementarity between NIKA2 observations of SZ effect and the follow-up that could be done with interferometers.

\begin{figure}[h!]
	\centering
	\begin{minipage}[b]{0.45\linewidth}
		\includegraphics[width=1.0\textwidth]{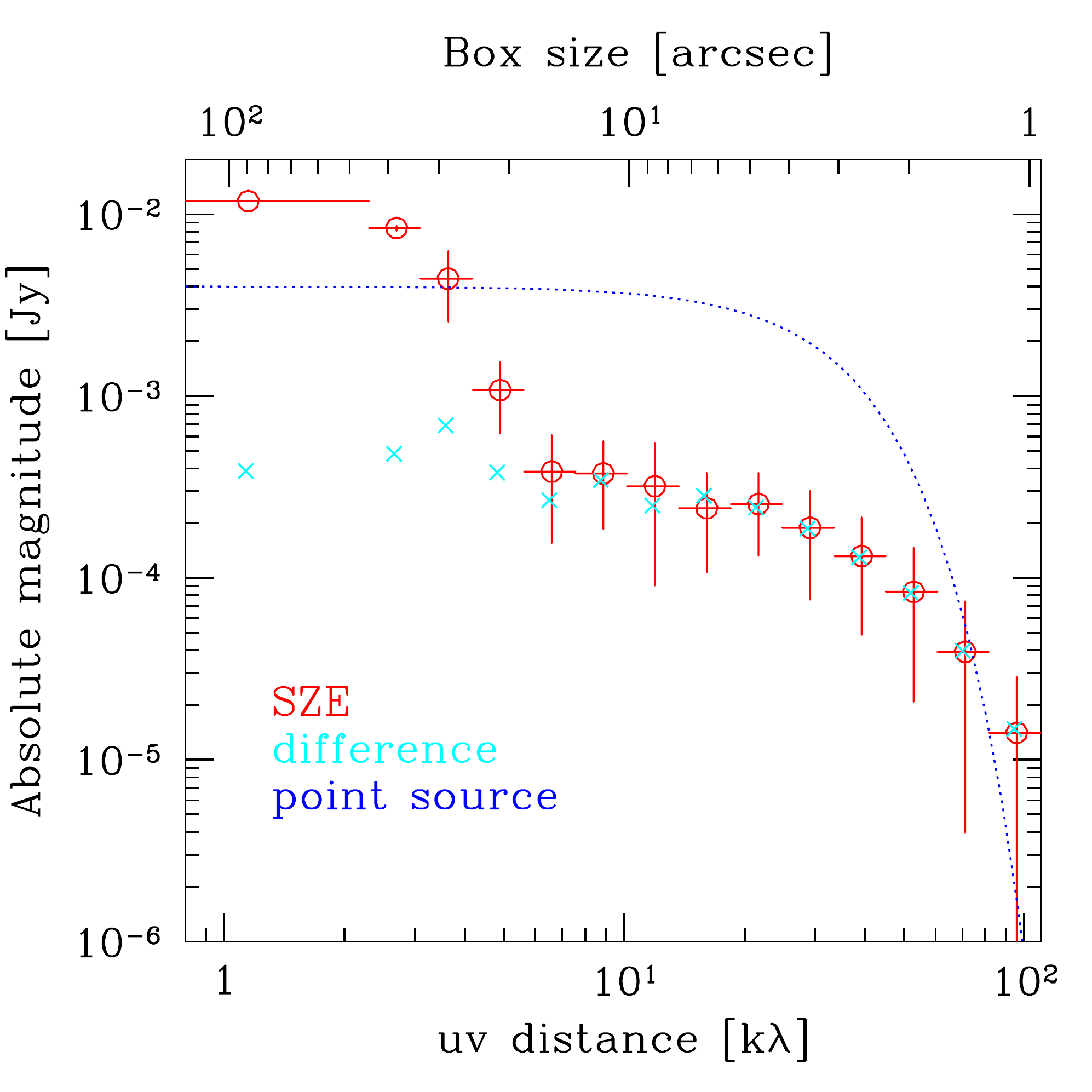}
	\end{minipage}
	\begin{minipage}[b]{0.45\linewidth}
		\includegraphics[width=1.3\textwidth]{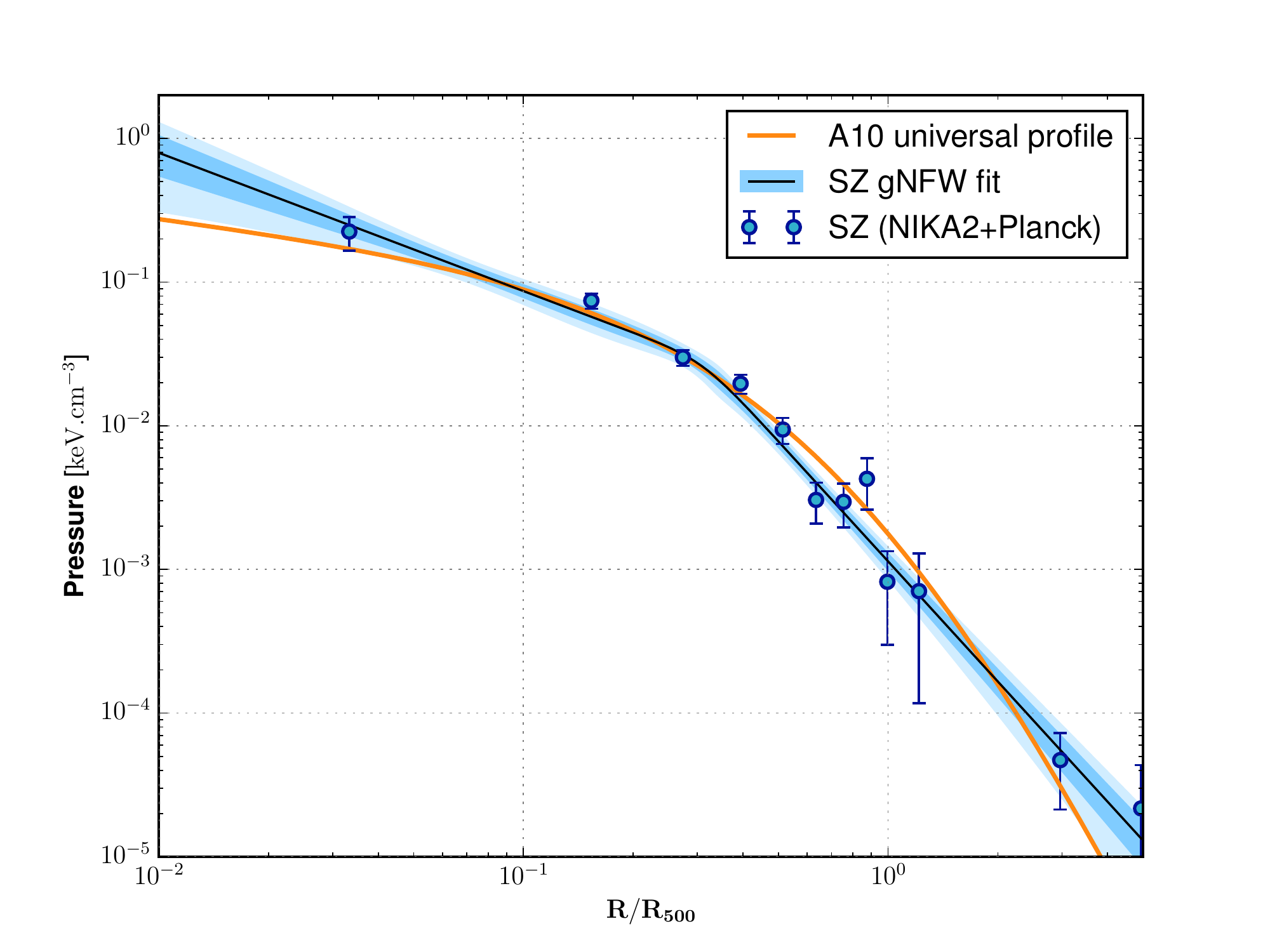}
	\end{minipage}
	\caption{{Left:} SZ effect signature of RX J1347.5–1145 {at 92\,GHz} imaged by ALMA after deconvolution as shown by red open circles with their associated error bars \cite{2016Kitayama}, a Gaussian source of 4\,mJy/beam with a size of 3.2\arcsec (FWHM) is also displayed in dashed blue. Top x-axis show that going up to 20 to 40\arcsec scales is suitable to retrieve the SZ signature in the Fourier domain. {Right:} Pressure profile of PSZ2 G144.83+25.11 (in blue) constrained by NIKA2 (blue circles) and Planck \cite{2018Ruppin} which is compared to the Universal Pressure Profile \cite{2010Arnaud} with an orange line. NIKA2 data and associated fit suggest a departure from the Universal Profile at small radii.}
	\label{fig-3}       % Give a unique label
\end{figure}

\noindent SZ with interferometers present some advantages with respect to single dish imaging. In particular, with interferometry it is possible to disentangle  the diffuse negative SZ emission from the small-scale positive point source contamination thanks to spatial filtering. Indeed, SZ effect is primarily present in short baselines only which is not the case of point sources (see Fig.~\ref{fig-3}, left panel). Interferometry also {implies} an accurate beam knowledge, and rejection of some systematics thanks to baseline correlations. The high angular resolution of interferometers make them sensitive to small scale temperature gradients and suitable to characterize pressure substructures in the ICM up to scales $\sim$ $\lambda/2D$ (see Fig.~\ref{fig-3}, left panel). While the shortest NOEMA projected baseline is 15\,m in classic mapping mode, it can be lowered down to 7.5\,m by performing a mosaic \cite{1979Ekers,2010Pety} allowing to fit the SZ signature in the \textit{uv} plane up to $\sim$\,28\arcsec scales at 2\,mm and $\sim$\,58\arcsec at lower edge of the 3\,mm band of NOEMA (72\,GHz). The new capability of NOEMA are promising to make joint studies of SZ effect between NOEMA and NIKA2. For example, this kind of joint study was made with ALMA, BOLOCAM and Planck to propose an alternative scenario to the strong shock-induced pressure perturbation deduced from X rays observations \cite{2019DiMascolo}. On the other hand, the ALMA SZ image of RX J1347.5–1145 was in good agreement with the electron pressure map built independently from the X-ray data \cite{2016Kitayama}. While NIKA2 {LP SZ} will allow to establish a Universal Pressure Profile at high redshift and study its evolution, NOEMA could help to investigate a sub-sample with indications of overpressure like departure from the current Universal Profile \cite{2010Arnaud} at small radii (see Fig.~\ref{fig-3}, right panel).

%
% BibTeX or Biber users please use (the style is already called in the class, ensure that the "woc.bst" style is in your local directory)
% \bibliography{name or your bibliography database}

\end{document}